# Calibration of a Digital Hadron Calorimeter with Muons

Burak Bilki[d], John Butler[b], Tim Cundiff[a], Gary Drake[a], William Haberichter[a], Eric Hazen[b], Jim Hoff[c], Scott Holm[c], Andrew Kreps[a], Ed May[a], Georgios Mavromanolakis[c,1], Edwin Norbeck[d], David Northacker[d], Yasar Onel[d], José Repond[a], David Underwood[a], Shouxiang Wu[b], Lei Xia[a]

[a]*Argonne National Laboratory, 9700 S. Cass Avenue, Argonne, IL 60439, U.S.A.*
[b]*Boston University, 590 Commonwealth Avenue, Boston, MA 02215, U.S.A.*
[c]*Fermilab, P.O. Box 500, Batavia, IL 60510-0500, U.S.A.*
[d]*University of Iowa, Iowa City, IA 52242-1479, U.S.A.*

**Abstract.** The calibration procedure of a finely granulated digital hadron calorimeter with Resistive Plate Chambers as the active elements is described. Results obtained with a stack of nine layers exposed to muons from the Fermilab test beam are presented.

**Keywords:** Calorimetry, Linear Collider, Particle Flow Algorithms, Resistive Plate Chambers.

**PACS:** 29.40.Vj, 29.40.Cs, 29.40.Gx

## INTRODUCTION

Particle Flow Algorithms (PFAs) attempt to measure all particles (originating from the interaction point of a typical colliding beams detector) in a jet individually, using the detector component providing the best momentum/energy resolution. The momenta of charged particles are measured with the tracking system (except for high momenta, where the calorimeter provides a better measurement), the energy of photons are measured with the electromagnetic calorimeter (ECAL), and the energy of neutral hadrons, i.e. neutrons and $K_L^0$'s, are measured with both the ECAL and the hadronic calorimeter (HCAL). The energy of a jet is reconstructed by adding up the energy of the individual particles identified as belonging to the jet. The major challenge in this approach to the measurement of jet energies lies in the identification of energy deposits in the calorimeter belonging to either a charged or neutral particle. Hence the requirement of calorimeters with very fine segmentation of the readout. Additional details on PFAs and the requirements for calorimetry can be found in references [1,2].

---

[1] Also affiliated with University of Cambridge, Cavendish Laboratory, Cambridge CB3 OHE, U.K.

In this context this paper reports on the development of a finely granulated HCAL using Resistive Plate Chambers (RPCs) as the active medium, which meets the requirements of PFAs. In preparation for the construction of a larger prototype module, a stack of nine small chambers was assembled and exposed to the muons, electrons and pions of the Fermilab test beam. Following the description of a general calibration procedure for such calorimeters, the measurements performed with the broad-band muon beam are described in detail.

## DESCRIPTION OF THE CALORIMETER STACK

The calorimeter stack consisted of nine chambers interleaved with the combination of a steel (16 mm) and a copper (4 mm) absorber plates, corresponding to approximately 1.2 radiation length. The chambers measured 20 x 20 cm$^2$ and were based on two separate designs. Eight chambers consisted of two glass plates (default design), whereas one chamber featured only one glass plate (so-called 'exotic' design). A schematic of the two chamber designs is shown in Figs. 1 and 2. The thickness of the glass plates was 1.1 mm and the gas gap was maintained with fishing lines with a diameter of 1.2 mm. The overall thickness of the chambers, including layers of Mylar for high voltage protection, was approximately 3.7 mm.

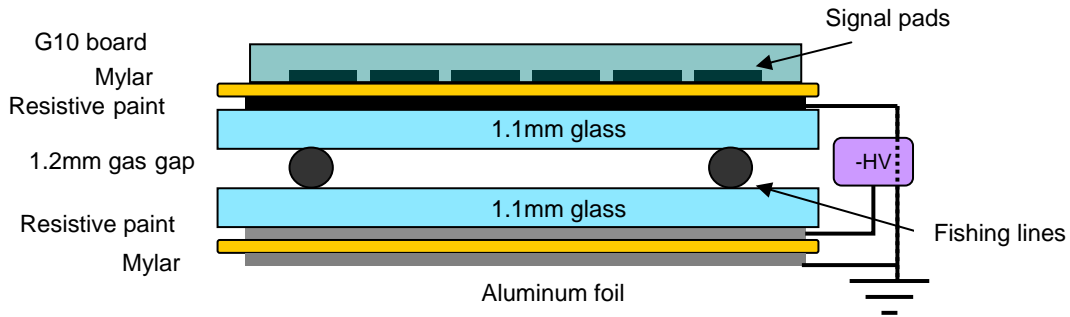

Figure 1. Schematic of the default chamber design with two glass plates. Not to scale.

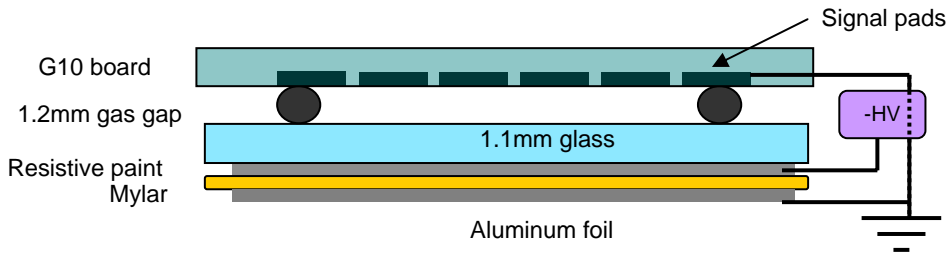

Figure 2. Schematic of the 'exotic' chamber design with one glass plate. Not to scale.

The chambers were operated in saturated avalanche mode with an average high voltage setting around 6.1 kV. The gas consisted of a mixture of three components: R134A (94.5%), isobutane (5.0%) and sulfur-hexafluoride (0.5%). For more details on the design and performance of the chambers, see [3].

The chambers were mounted on the absorber plates and these in turn were inserted into a hanging file structure. The gap between absorber plates was 13.4 mm, where 8.3 mm were taken by the chambers and their readout boards.

## DESCRIPTION OF THE ELECTRONIC READOUT SYSTEM

The electronic readout system is optimized for the readout of large numbers of channels. In order to avoid an unnecessary complexity of the system, the charge resolution of individual pads is reduced to a single bit (digital readout). The system consists of several parts which are briefly described in the following:

a) Pad-board: the pad board is a four-layer board containing the signal pick-up pads on one (the RPC) side and gluing pads on the other side. The signals are transferred from one to the other side via transfer lines. The dimensions of the board are 20 x 20 cm$^2$, where the central 16 x 16 cm$^2$ contain 256 1 x 1 cm$^2$ signal pick-up pads.
b) Front-end board: the front-end board contains the mirror-image of the pad-board's gluing pads on one side and four front-end Application Specific Integrated Circuits (ASICs) on the other side. Particular care was devoted to the shielding of the analog signal lines from the digital lines of the ASIC. Conductive adhesive was used to connect the gluing pads of the pad- and front-end boards.
c) DCAL chip: the DCAL chip served as front-end ASIC, reading out 64 individual pads. The chip can be operated in high or low gain mode, where the latter is more suited for operating with RPCs. The threshold value is common to all channels of a chip and is set by an internal digital-to-analog converter (DAC) with a range of 256 counts. In low gain mode, the DAC range corresponds to threshold settings between ~20 and 700 fC. The output of the chip is a hit pattern (64 bits) and a time-stamp with a resolution of 100 ns. The chip can be operated in either triggered or self-trigger mode.
d) Data concentrator: each data concentrator receives data from four front-end ASICs. Events with zero hits in a given ASIC are suppressed at this stage.
e) Data collector: the VME-based data collectors receive data from the front-end and format them for transfers to the data acquisition computer. A commercial VME-PCI bridge served as link to the computer. Slow control signals for device configuration are sent from the data collector via the data concentrator to the front-end ASICs.
f) Trigger and timing module: a VME-based trigger and timing module generates the clock and time-stamp reset signals for the readout and distributes the external trigger signals to the data collector modules.

The total number of readout channels was up to 2,304, for nine layers. More details, in particular, on certain performance aspects of the electronic readout system can be obtained from [4].

# CALIBRATION PROCEDURE

In a digital hadron calorimeter the energy of an incoming hadron, $E_{hadron}$, can be reconstructed from the number of hits associated with that particle. Ignoring effects of high-density sub-clusters, which might require non-linear corrections, Eqn. 1 provides a general formula for reconstructing $E_{hadron}$

$$E_{hadron} = \alpha_{samp}(\Sigma_i H_i) \cdot \Sigma_i (H_i - B_i)/(\varepsilon_i^{MIP} \cdot \mu_i^{MIP}), \qquad (1)$$

where

- $i$ is an index running over all pads associated with an incoming particle
- $H_i$ is set to 1 (0) depending on whether a hit (or no hit) has been recorded in pad $i$
- $B_i$ are the expected number of background hits (from accidental discharges, electronic noise or cosmic rays) for pad $i$
- $\varepsilon_i^{MIP}$ is the efficiency for pad $i$ to fire when traversed by a minimum ionizing particle (MIP)
- $\mu_i^{MIP}$ is the average number of pads firing when pad $i$ is traversed by a MIP and
- $\alpha_{samp}(\Sigma_i H_i)$ is a sampling term, possibly depending on the total number of hits.

The sampling term can be regarded as independent of specific running conditions and so to be time independent. It needs to be determined from test beam data and Monte Carlo simulations.

The chamber's calibration constants ($B$, $\varepsilon^{MIP}$, $\mu^{MIP}$) depend on both operational (such as high voltage and threshold settings) and environmental conditions (such as temperature, humidity and atmospheric pressure). In the following we study the dependence on the operational parameters, assuming that the environmental effects are small in comparison. Thus no corrections are applied for the environmental dependencies. Operation of the chambers in saturated avalanche mode results in an equal response from particles with different masses (pions, protons...), such that the calibration constants, as established with muons, can be generalized to all particles in a hadronic shower.

In a colliding beam experiment, the calibration of a RPC fine-grained digital hadron calorimeter can be maintained without the need of a dedicated calibration system, as any track segment within a hadronic shower can be used to obtain a measurement of the efficiency and pad multiplicity.

# MEASUREMENT OF THE NOISE RATE

The noise rate was measured using the self-trigger data acquisition mode of the DCAL chip. For a given high voltage setting, Fig. 3 shows the noise rate as a function of threshold for a selection of chambers. Two chambers, with an overall noise rate a factor 5 – 10 higher, due to hot spots, were omitted from the plot. The rate at our default threshold setting of 110 DAC counts corresponds to about 0.15 Hz/cm$^2$. Given a 300 ns gate width for triggered data acquisition, this rate translates into an accidental rate of $4.5 \cdot 10^{-8}$ hits/pad/event. Extrapolated to a calorimeter with say $50 \cdot 10^6$ channels, as envisaged for the International Linear Collider (ILC), this rate in turn corresponds to 0.2 hits/event. With the high voltage to the chambers turned off the noise rate was found to be less than $4 \cdot 10^{-5}$ Hz/cm$^2$.

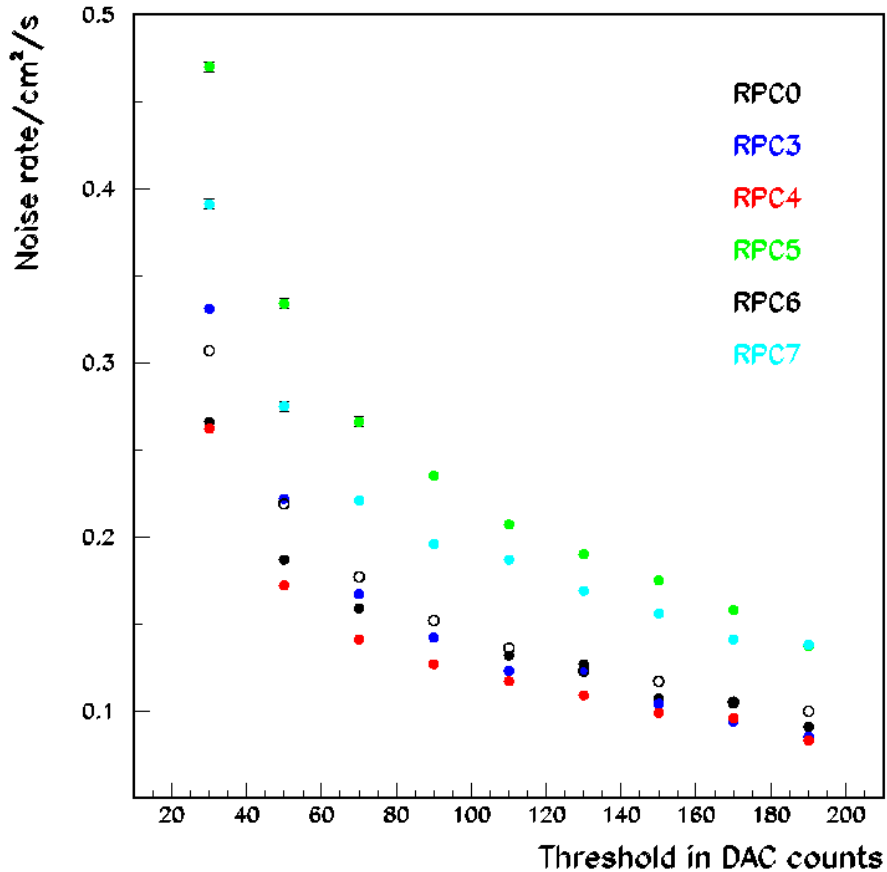

Figure 3. Rate of noise hits per readout pad as a function of threshold for 6 different chambers. The measurements were performed with a high voltage setting of 6.3 (6.0) kV, for the default (exotic) chambers. The open circles indicate the results of the 'exotic' chamber (RPC6)

For a threshold setting of 110 counts, Fig. 4 shows the dependence of the noise rate on the high voltage of the chamber. The noise rate is seen to increase approximately quadraticaly with high voltage.

Figure 5 shows the x – y map of the noise hits for a typical default and the 'exotic' chamber. A clear clustering along the fishing lines located approximately at x = 4.2 and 10.7 is visible. The noise rate was observed to fluctuate significantly over longer periods of time in comparison to the 5 minute noise data collection runs. Further studies to establish the impact of the environmental variables on these rates are ongoing.

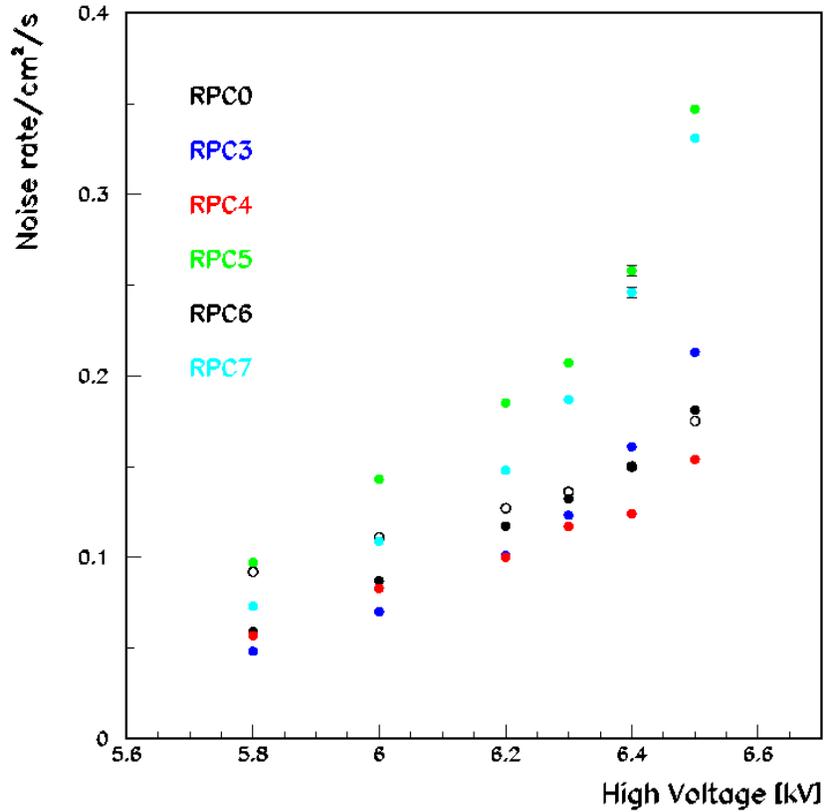

Figure 4. Rate of noise hits per readout pad as a function of high voltage for 6 different chambers. The measurements were performed with a threshold at 110 counts. The high voltage values for the exotic chamber (open circles) were set at 300 V less than indicated in the plot.

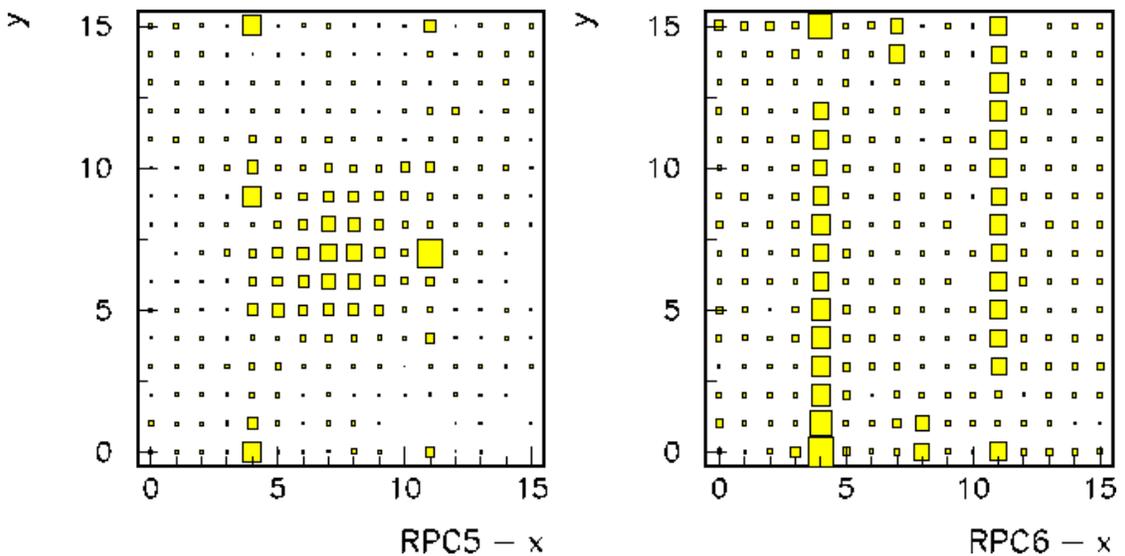

Figure 5. x-y map of noise hits for a default and the exotic chambers, where the rate is proportional to the area of the shaded boxes. The fishing lines are located at x = 4.2 and 10.7 cm.

# TEST BEAM SETUP AND DATA COLLECTION

The calibration procedure was performed at the Meson Test Beam Facility (MTBF) of Fermilab [5]. The calibration runs were taken with the primary 120 GeV proton beam and a 9 foot (~ 3 m) iron beam blocker in place to produce muons and filter out the non-muon component. The readout of the stack was triggered by the coincidence of two large scintillator paddles, each with an area of 19 x 19 $cm^2$, located approximately 5.0 and 0.5 meters upstream of the stack. A photograph of the setup is shown in Fig.6.

The beam came in spills of four second length every one minute. The trigger rates were typically between 100 and 200 per spill. At these low rates the long recharge times (of the order of 1 msec) of the RPCs is not expected to decrease the efficiency of the chambers.

For a given run, all default chambers were operated at the same high voltage and threshold settings. The 'exotic' chamber was operated at the same threshold, but at a somewhat lower voltage, to ensure a similar MIP detection efficiency to the one obtained with the default chambers. Table I summarizes the various operating conditions. For each setting 5,000 – 10,000 events were collected in separate data acquisition runs.

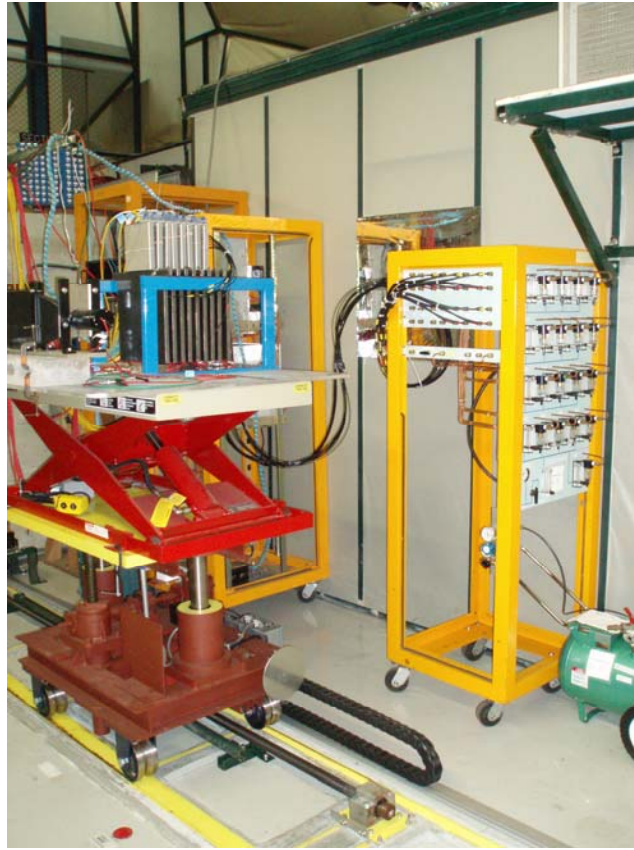

Figure 6. Photograph of the setup at the Meson Test Beam Facility of Fermilab. The stack containing nine layers within the blue hanging file structure and part of the closer trigger scintillation counter are visible on the jack stand at left. The gas distribution rack is shown to the right.

From these data sets, Fig. 7 shows various views/projections of two selected events. Figure 7.a shows the event display of a clean muon track, where in every layer a single pad fired. Depending on the location of the avalanche with respect to the pad boundaries up to four pads might fire. The event display of Fig. 7.b shows a muon track entering the calorimeter with a slight inclination both horizontally and vertically. As a result some layers show more than one hit in both projections.

| Number of Chambers in the stack | High Voltage in kV | Threshold in DAC counts |
|---|---|---|
| 8 | 6.2/5.9 | 30 |
|   |   | 50 |
|   |   | 70 |
| 9 | 6.3/6.0 | 30 |
|   |   | 70 |
|   |   | 110 |
|   |   | 150 |
|   |   | 210 |
| 7 | 6.4/5.8 | 30 |
|   |   | 50 |
|   |   | 70 |
|   |   | 110 |
|   |   | 150 |
|   |   | 190 |
|   |   | 210 |
| 8 | 6.5/6/2 | 30 |
|   |   | 120 |
|   |   | 210 |

Table I. Summary of calibration runs. The first/second number for the high voltage corresponds to the default/exotic chamber settings. For each setting between 5,000 and 10,000 events were collected.

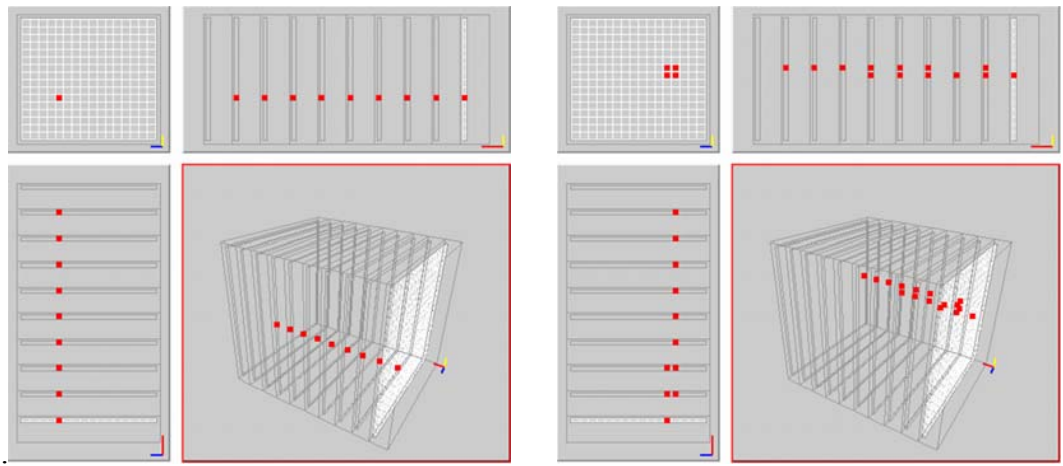

Figure 7. Event displays of single muon tracks: a) with one hit per layer, and b) entering the calorimeter at a slight angle and therefore leading to multiple hits in some layers.

# ANALYSIS PROCEDURE

The analysis of the test beam data selects clean single muon track segments for the measurement of the MIP detection efficiency and the pad multiplicities. Due to the loose trigger conditions, some selection criteria for rejecting multiple or showering particles were necessary. The analysis involved several steps, which are described in detail in the following:

a) *Rejection of events with large number of hits.* In order to reduce the contamination from multiple or showering particles, events with more than 50 hits were rejected, where a 'hit' is defined as a single pad firing, possibly as part of a cluster of nearby pads. Typical single muon induced events have between 1 or 2 hits per layer.
b) *Clustering of the hits in each layer.* As a first step all hits in a given layer are assigned into clusters of touching cells, where 'touching' is defined as sharing a common side. The center-of-gravity of a cluster i ($x_i$, $y_i$) is obtained as the average position of the hits.
c) *Reconstruction of track segments.* In order to reconstruct track segments, only clusters with at most four hits are being considered. This is to avoid a measurement bias due to multiple tracks or showering particles. For the first (last) layer in the stack, clusters in the following (preceding) two layers are being considered for track reconstruction. For intermediate layers, clusters in the layer immediately preceding and in the layer just following are being considered. A track segment is defined by two aligned clusters in these layers, where the difference in center-of-gravity of the clusters $\Delta R^2 = \Delta x^2 + \Delta y^2 < 9$ cm$^2$.
d) *Fiducial cuts to avoid the readout boundaries.* Once aligned clusters have been identified, the track through their centers-of-gravity is extrapolated (or interpolated for intermediate layers) onto the layer being investigated. If the extra/interpolated track position is within 1 cm of the border of the readout area, the track is discarded. This fiducial cut ensures a proper measurement of the pad multiplicity and avoids a bias due to the limited area of the readout boards.
e) *Fiducial cut associated with readout problems.* During the data taking period either all or part of the 64 channels of some DCAL chips developed readout problems. Since these problems were not correlated to the chamber performance, a fiducial cut around the dead area of the chips, including a one-pad wide rim around the area, was introduced. This problem affected approximately 5% of the readout channels.[2]
f) *Additional cuts.* In order to reduce the effect from showering particles, tracks are eliminated if any layer, apart from the layer being investigated, counts more than 10 hits. This cut introduced a small systematic uncertainty of 0.4% (0.04) on the efficiency (pad multiplicity).

---

[2] This problem was later attributed to the particular grounding scheme of the test beam setup. An improved grounding eliminated all these problems.

g) *Matching with clusters in the layer being investigated.* In the layer being investigated clusters are searched for in the area around the extra/interpolated track position. Any cluster with a $\Delta R^2 < 10$ cm$^2$ from the extra/interpolated position is considered a match. The number of hits associated to this cluster enters the measurement of the pad multiplicity. If no cluster with $\Delta R^2 < 10$ cm$^2$ is found, the layer is considered inefficient in this event.

An independent analysis based on track reconstruction using all RPC layers (i.e. a global fit), rather than using track segments reconstructed in pairs of chambers, was also performed. The results of the two approaches were found to be very consistent.

## DEPENDENCE ON THRESHOLD

This section presents the dependence of the chamber characteristics (efficiency and pad multiplicity) as a function of the signal threshold. The results were obtained while operating the default chambers at 6.3 kV. Due to the smaller distance between anode and cathode plane, the 'exotic' chamber was operated at 6.0 kV, which resulted in similar MIP detection efficiencies. The beam block produced a beam spot size which illuminated evenly the chamber's readout area.

Figure 8 shows the distributions of number of hits as seen in the nine chambers for a threshold setting of 110 counts. The distributions for the individual chambers are quite similar. Only RPC7 (based on the 'exotic' design) shows a significantly different behavior from all others. Due to the smaller distance of the signal pads to the gas volume, clusters only rarely contain more than a single pad.

For a given chamber i the efficiency $\varepsilon_i^{MIP}$ for detecting a MIP is defined as the ratio of the number of track segments with at least one hit over the total number of track segments

$$\varepsilon_i = N_i^{hits>0} / N_i^{total}. \qquad (2)$$

Measurements of the efficiency as a function of threshold are shown in Fig. 9. At a threshold of 30 counts the chamber efficiencies are around 95%. With higher thresholds the efficiencies gradually decrease.

For a given chamber i the pad multiplicity $\mu_i^{MIP}$ is defined as the average number of hits per track segment, where the 'zeros' are excluded from the average. Figure 10 shows the average pad multiplicity versus threshold for the nine chambers individually. Whereas the default chambers show a decrease of the pad multiplicity with increasing threshold, the values for the 'exotic' chamber remain constant at around 1.1.

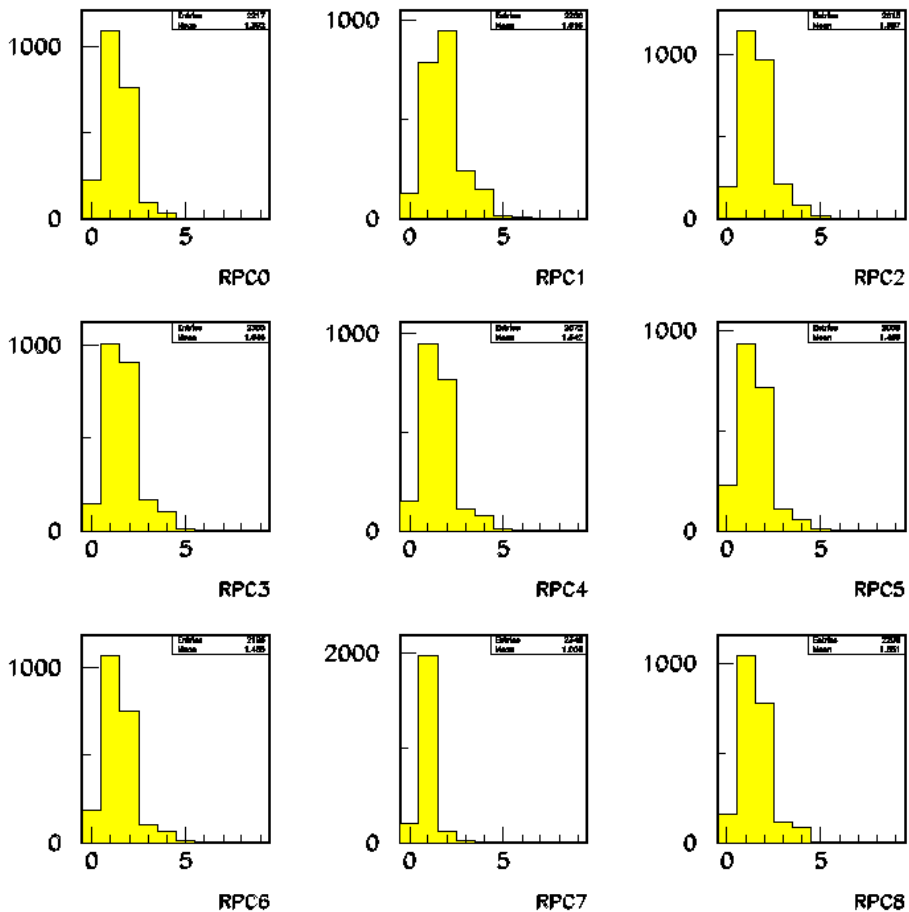

Figure 8. Distribution of hits per track segment for RPC0 through RPC8. The chambers were operated with a high voltage of 6.3/6.0 kV. The threshold was set at 110 counts. Note that RPC7 is the 'exotic' design chamber.

RPC0 (RPC1) shows a lower (higher) efficiency together with a lower (higher) pad multiplicity. This suggests that the operating voltage for these chambers should have been chosen somewhat differently to obtain similar results as for the other default chambers: higher for RPC0 and lower for RPC1 (see the next section for the high voltage dependence). RPC5 shows a lower efficiency together with average numbers of the pad multiplicity. This is due to a lower efficiency in one corner of the chamber, the cause of which was later related to the particular grounding scheme employed during the beam tests.

Figure 11 shows the pad multiplicity as a function of efficiency for all nine chambers. The results appear to lie on a common curve, even for the chambers RPC0 and RPC1 which require a somewhat different high voltage setting. Due to its efficiency problems, the results for RPC5 for a given pad multiplicity are shifted towards lower values for the efficiency. Again, the 'exotic' chamber shows a constant pad multiplicity over a wide range of efficiencies.

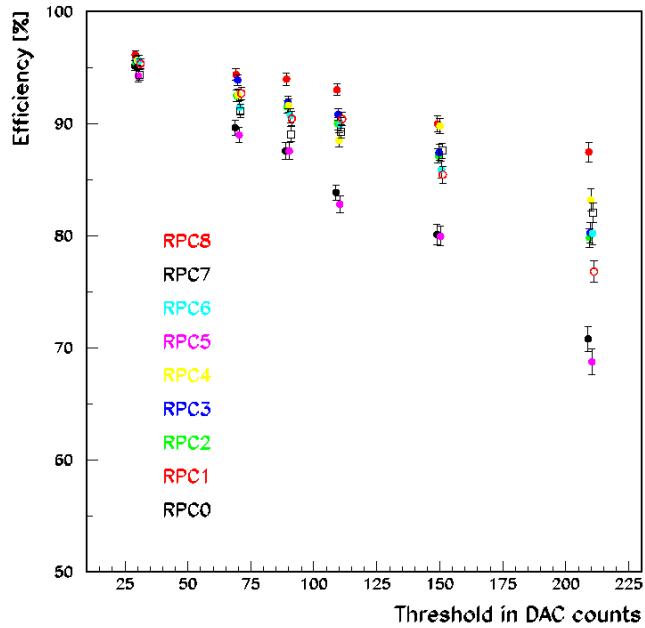

Figure 9. MIP detection efficiencies as function of threshold. The results of the 'exotic' chamber are shown as open squares. The measurements were performed with a high voltage setting of 6.3/6.0 kV. The results for RPC0-8 are shown individually.

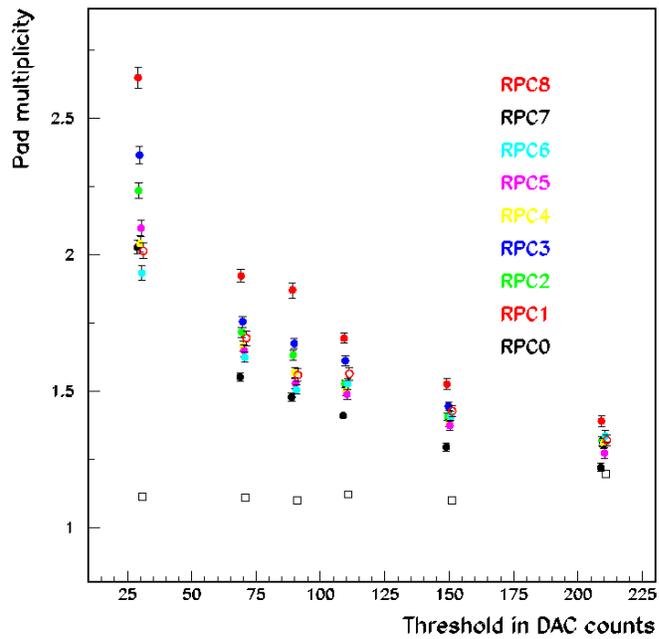

Figure 10. Pad multiplicity as a function of threshold. The results of the 'exotic' chamber are shown as open squares. The measurements were performed under the same conditions as for Fig.9.

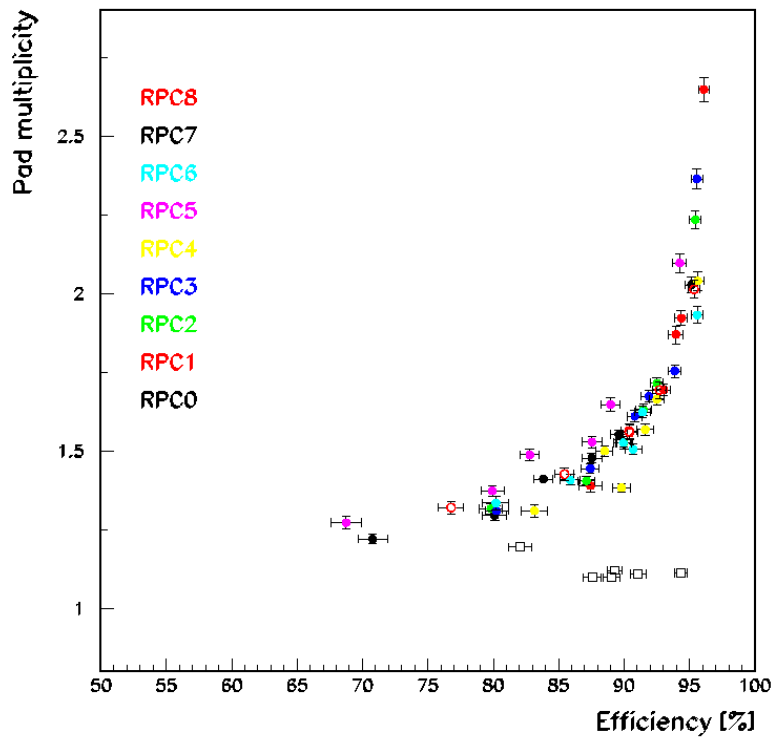

Figure 11. Pad multiplicity versus MIP detection efficiency for RPC0 through RPC8. The results of the 'exotic' chamber are shown as open squares. The measurements were performed under the same conditions as for Fig. 9.

# DEPENDENCE ON HIGH VOLTAGE

In order to study the dependence on the high voltage setting, the individual results for the various chambers have been averaged using weights inversely proportional to the measurement errors. Four chambers with obvious differences in operational high voltage requirements or with areas of inefficiencies have been excluded from these averages. Due to its lower and near constant pad multiplicity, the exotic chamber has been excluded as well.

Figure 12 shows the MIP detection efficiency as a function of threshold for four different high voltage settings. For a given threshold a clear increase in efficiency with higher voltage is observed.

The average pad multiplicity as a function of threshold is shown in Fig. 13. Again operation at higher voltages results in higher pad multiplicities.

Figure 14 shows the average pad multiplicities versus average efficiencies for four different high voltage settings. For an efficiency of 95% (90%) average pad multiplicities between 1.7 and 2.1 (1.4 and 1.6) are obtained. Furthermore, for a given efficiency, operation at higher voltages is seen to result in somewhat lower pad multiplicities. This effect was not observed in previous tests with an analog readout system [3]. The details of this discrepancy are still under investigation.

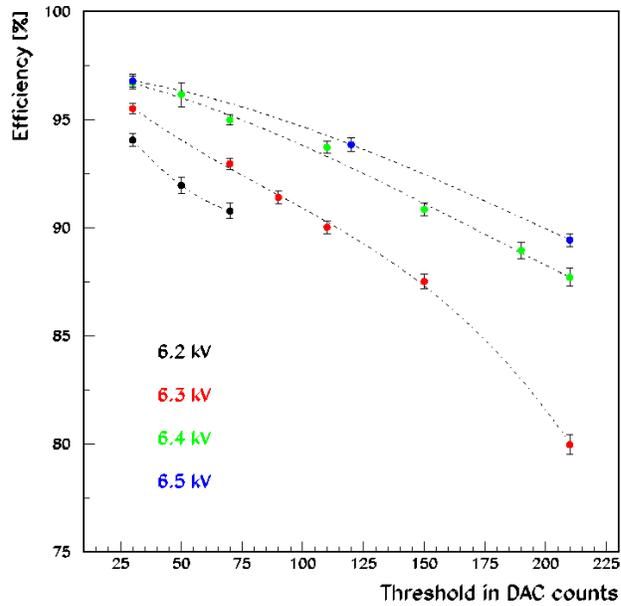

Figure 12. Average MIP detection efficiencies as a function of threshold for four different high voltage settings. Only chambers with similar operating points and no areas of inefficiencies have been included in this measurement. The lines are fit to higher order polynomials and only serve to guide the eye.

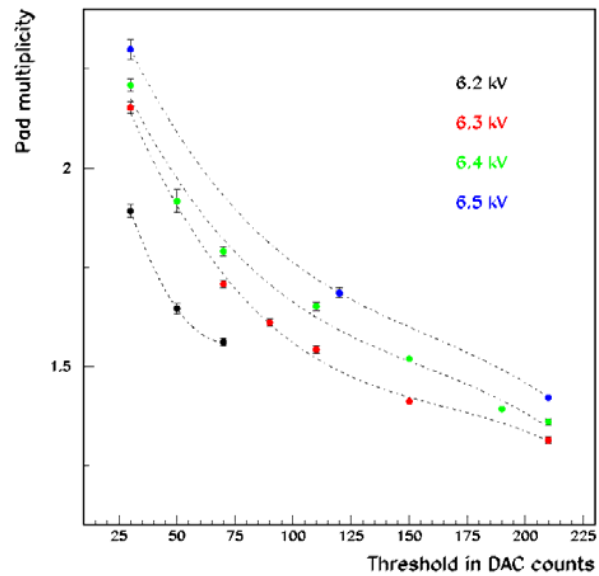

Figure 13. Average pad multiplicities as a function of threshold for four different high voltage settings. Only chambers with similar operating points and no areas of inefficiencies have been included in this measurement. The lines are fit to higher order polynomials and only serve to guide the eye.

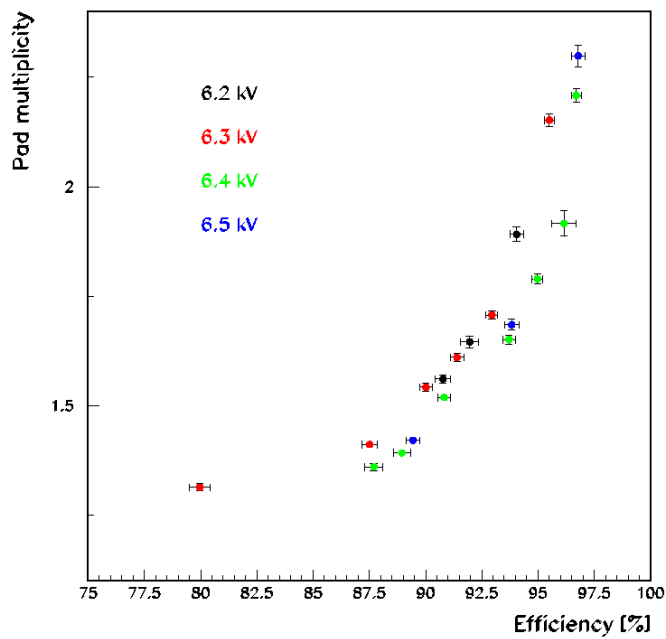

Figure 14. Average pad multiplicity versus average efficiency for four different high voltage settings. Only chambers with similar operating points and no areas of inefficiencies have been included in this measurement.

## X – Y MAP

Using only those chambers with no obvious areas of inefficiencies, Fig. 15 shows an x – y map of the inefficiency (defined as $1 - \varepsilon$) as averaged over the chambers. The data were collected with a high voltage of 6.3 kV and a threshold of 110 counts with an overall MIP detection efficiency of approximately 90%. In this figure, the largest boxes correspond to values of the inefficiency of approximately 15%. Two areas of higher inefficiencies around x = 4 and 11 are apparent. These areas coincide with the location of the fishing lines.

## CONCLUSIONS

A stack of up to nine layers of Resistive Plate Chambers (RPCs), interleaved with copper and steel absorber plates was exposed to a broad-band muon beam at the Fermilab Meson Test Beam Facility (MTBF). The data were used to establish the calibration parameters of such a calorimeter.

The calibration of a digital hadron calorimeter depends on a geometric sampling term (to be determined with test beam measurements and Monte Carlo simulations) and three possibly time-dependent quantities that are functions of high voltage, threshold and environmental conditions:

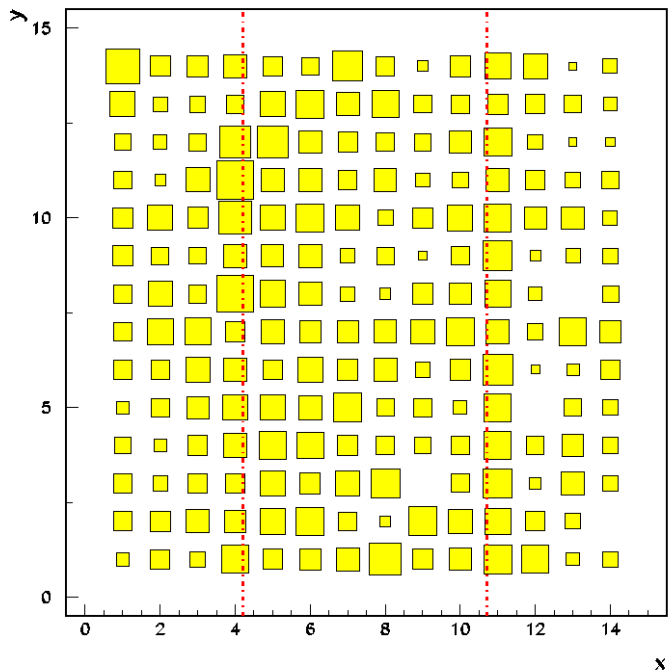

Figure 15. x-y map of the inefficiency as obtained by averaging over all chambers with no obvious areas of inefficiency. The largest boxes correspond to inefficiencies of about 15%. The fishing lines, used to keep the glass plates apart, are located at approximately x = 4.2 and 10.7 and are shown as dashed lines.

i) Background noise: Using the self-triggered mode of the front-end readout, the rate was established to be typically 0.15 Hz/cm$^2$. The probability of such a noise hit overlapping with a 300 ns readout window, as used in triggered mode, is negligible.

ii) MIP detection efficiency: Depending on the high voltage and threshold settings efficiencies in the range between 80% and 96% were obtained using track segments reconstructed in neighboring layers. As expected the efficiency drops around the location of the two fishing lines located in the gas volume. Lower values of the efficiency observed in two of the chambers were later explained as being related to the particular grounding scheme used during the test beam data taking.

iii) Pad multiplicities: Depending on the high voltage and threshold settings pad multiplicities between 1.2 and 2.2 were measured using track segments in neighboring layers. With the 'exotic' chamber pad multiplicities around 1.1 were obtained, independent of the operational conditions.

For a given high voltage, the pad multiplicities versus MIP detection efficiencies lie on a common curve, despite slight differences in high voltage requirements between individual chambers. Operation at a higher high voltage results in a relatively larger gain in efficiency than in pad multiplicity.

Enhanced losses of efficiency are observed at the location of the fishing lines in the gas volume.

Overall, the 'exotic' compared to the default RPC design offers two distinct advantages: a lower average pad multiplicity and a reduced height. During the test beam runs at Fermilab the chamber performed reliably with an overall noise rate comparable to the other default design chambers. However, it remains to be seen if long-term effects, e.g. due to the exposure of the pads to the gas mixture, will limit its usefulness in detectors with an expected lifespan of a decade or more.

# ACKNOWLEDGEMENTS

We would like to thank the Fermilab test beam crew, in particular Erik Ramberg, Doug Jensen, Rick Coleman and Chuck Brown, for providing us with excellent beam. The University of Texas at Arlington ILC group is acknowledged for providing the two trigger scintillator paddles and their associated trigger logic.